\documentclass[a4paper,12pt,reqno,superscriptaddress,nofootinbib]{revtex4}
\usepackage[centertags]{amsmath}
\usepackage{amsfonts}
\usepackage{amssymb}
\usepackage{amsthm}
\usepackage{newlfont}
\usepackage{stmaryrd}
\usepackage{mathrsfs}
\usepackage{mathtools}
\usepackage{euscript}
\usepackage{graphicx}
\usepackage{enumerate}
\usepackage{todonotes}

\usepackage{color}

\usepackage{tikz}
\usepackage{pgf}
\usetikzlibrary{positioning,fit,calc}
\usetikzlibrary{arrows,automata}
\usepackage{wrapfig}
\usepackage{subfigure}
\usepackage{amscd}
\usepackage{hyperref}

\usepackage{changes}

\theoremstyle{plain}

\theoremstyle{definition}

\theoremstyle{remark}

\let\ve=\varepsilon

\newcommand{\be}{\begin{equation}}
\newcommand{\en}{\end{equation}}

\newcommand{\opunit}{\text{1}\kern-0.22em\text{l}}

\newcommand{\id}{\textrm{d}}

\DeclareMathAlphabet{\mathpzc}{OT1}{pzc}{m}{it}

\let\oldsqrt\sqrt
\def\sqrt{\mathpalette\DHLhksqrt}
\def\DHLhksqrt#1#2{%
	\setbox0=\hbox{$#1\oldsqrt{#2\,}$}\dimen0=\ht0
	\advance\dimen0-0.2\ht0
	\setbox2=\hbox{\vrule height\ht0 depth -\dimen0}%
	{\box0\lower0.4pt\box2}}

\let\ve=\varepsilon

\let\be=\beta

\DeclareMathAlphabet{\mathpzc}{OT1}{pzc}{m}{it}

\def\bea{\begin{eqnarray}}
\def\eea{\end{eqnarray}}
\def\ba{\begin{array}}
	\def\ea{\end{array}}

\usepackage{changes}

\let\ve=\varepsilon
\begin{document}

\title{Pushing run-and-tumble particles through a rugged channel}
\author{Bram Bijnens and Christian Maes 
	\\ {\it Instituut voor Theoretische Fysica, KU Leuven}}

\email{christian.maes@kuleuven.be}


\begin{abstract}
	 We analyze the case of run-and-tumble particles pushed through a rugged channel both in the continuum and on the lattice.  The current characteristic is non-monotone in the external field with (1) the appearance of a current and nontrivial density profile even at zero field for asymmetric obstacles, (2) the current decreasing with persistence at small field and increasing with persistence at large field. Activity in terms of self-propulsion increases the maximal current and postpones dying. We give an effective theoretical description with wider validity.
\end{abstract}
\maketitle


\section{Introduction}
A nonequilibrium system, be it transient or stationary driven, is very sensitive to time-symmetric constraints and obstacles.   Kinetic constraints or disorder indeed influence relaxation, diffusion and transport even for independent particles. For transport in out-of-equilibrium systems, even at small (and even zero) external driving, the frenetic contribution complements entropic considerations (as in the fluctuation--dissipation relation) to enter crucially in the current characteristic \cite{fren,nondis}.  Examples where the response to external fields, temperature or chemical affinities show negative differential susceptibility include \cite{zia,neg,gar,oliver,sar,chemfalasco,rei1}. In the present paper we revisit the set up of \cite{zia,neg} but for active particles as studied before e.g. also in \cite{alo,patt,kret,rib,rei4,rei5}:  how does the current characteristic change as a function of the persistence?\\

Active matter consists of self-propelled particles, characterized by a coupling between motion and an internal degree of freedom, quantified by persistence \cite{ro,be,ra,fo,go}.  Their dynamics model bacteria-motion and nanomotors or active colloids for their locomotion, possibly showing collective effects such as flocking or phase separation \cite{to,ku,sch,re,ste,ca}.  Also individually they differ in many ways from their passive counterparts \cite{ma,ba,one}.  
The models are characterized as overdamped motion, while being nonMarkovian in the position variable by the presence of colored noise.\\
Pushing active particles can be understood in various ways.  We use the more neutral words of ``pumping'' or ``pushing'' to include janus-particles or active colloids but there are many specific ways in which also living organisms may respond to a signal and hence introduce bias into their movement \cite{revi}.  We think about a fixed sensory gradient where a direction is preferred and remains  the same for all individuals at all locations in space (e.g. from gravity). Or, when animals search for food, the direction may be taken of  a target or source fixed in space. \\
As a model for self-propelled colloids, we take run-and-tumble particles (RTP), mimicking a behavior observed in chemotactic bacteria \cite{co}.  We have no very specific biological application in mind however, and we can as well take active colloids as susceptible to a global external driving as today can be realized in many ways. For a related approach and type of problem, see e.g. \cite{ber} showing glassy behavior in an active model.  That is a form of kinetic arrest which is not unlike the trapping behavior we study in the present behavior, except that interactions will play no role here.  It is the roughness of the channel that creates possible trapping. \\
Recently, there have appeared a growing number of studies of active systems exposed to disorder and obstacles, with  \cite{chep} studying the influence of heterogeneity on large-scale collective properties, similarly for \cite{mor,qui} on flocking behavior and \cite{san,rei4,rei5} for active particles running in a landscape of obstacles, typically randomly-placed disks, and \cite{mor2,zei,bert,chepi,bhat,jak} investigating diffusion properties of self-propelled particles in crowded or complex environments. Periodic arrays of obstacles were considered for (artificial) microswimmers in \cite{volp} and transport  for active particles in periodic porous media was characterized in \cite{alo}, for periodic arrays (for active Brownian particles) in \cite{patt,kret,rib}.  In particular we mention \cite{rei3,rei2} for important new directions and  \cite{hao} for modifier activation--inhibition switching in enzyme kinetics. Non-Boltzmann steady distribution and clustering have been studied for RTP in \cite{dha,ta,sev,mal,be}.  New in the present paper is the study on the discrete lattice, incorporating the external field and activity at the same time, plus a general effective analysis able to summarize the properties of the rough substrate in terms of an effective mobility.\\

The plan is as follows:  in the next section, we specify the question, after which we present an effective dynamics to predict the main features of RTP-particles being pushed through a rough (but general) channel. Section \ref{mod} describes the model dynamics more specifically in the continuum and on the discrete lattice  through a channel with obstacles.  The obstacles lead (possibly) to trapping, as has been analyzed in various scenarios for passive particles.  In Section \ref{res} we present the results of simulations, where the main output is the current characteristic, function of the external field, and how it depends on persistence.  Section \ref{theo} concludes with the discussion, also with reference to the predictions in Section \ref{theor}.

\section{Question and effective analysis}\label{theor}

Active particles in a constant external field have been mostly considered for the problem of sedimentation.  The experimental system is for example treated by Palacci {\it et al.}, \cite{pal}.\\ 
Enculescu and Stark \cite{enc} considered active Brownian motion with an external time-independent constant force, studying sedimentation.  Tailleur and Cates \cite{tai} showed that for the run-and-tumble model the stationary
state distribution in a linear potential has the exponential form.  For that model, a closed form of the
stationary state distribution has been derived by Szamel in \cite{sza}.  In the present paper we investigate the current for RTP-motion driven through a rough periodic channel, where the two-dimensional aspect matters for the placement of obstacles.\\

In this Section, we restrict ourselves to continuous space.  (The aspects related to modeling RTP on the lattice in an external field will be given in Section \ref{dis}.)\\
Run-and-tumble particles run on straight lines interrupted by tumbling at exponential times. At tumbling a new random direction is chosen. In an external field $\ve$ pointing in the $x-$direction with unit vector $\boldsymbol{\hat{e}_x}$, the two-dimensional position $\vec{r}_t$ changes in time $t$ following the differential equation
\begin{equation}\label{eq: RTP in field}
\dot{\vec{r}}_t = c \,\hat{\sigma}_t + \ve\, \boldsymbol{\hat{e}_x}
\end{equation}
where $c>0$ is the amplitude of the noise $\hat{\sigma}_t$ which is itself a Markov process taking values in the space of unit-vectors (points on the circle):
\begin{equation}
\hat{\sigma_t} = \cos \theta_t\, \boldsymbol{\hat{e}_x} + \sin\theta_t\,\boldsymbol{\hat{e}_y}
\end{equation}

with  angle $\theta_t$ undergoing a jump process with constant rate $k(\theta,\theta') = a$.  In other words, at tumbling times, the particle randomly chooses a new angle to run.  The particle feels the direction and stimulus intensity $\ve$ at every single point  but its tumbling rate $a$ is unaffected.\\  We think of a channel, periodic along the $x-$direction, where the interior surface is not smooth. We ask for the stationary particle current and how it gets influenced by $c$ and $a$ and the channel properties. In Section \ref{mod} we will choose for hooks that block the motion. In general the dynamics \eqref{eq: RTP in field} must be supplemented with boundary conditions indeed, but next we assume an effective one-dimensional and translation-invariant model where the roughness is replaced by a mobility $\mu$.  \\

Consider then the effective one-dimensional dynamics 
\begin{equation}\label{1eff}
\dot{x}_t = \psi(c\sigma_t+\ve)
\end{equation} 
where $\sigma_t=\pm 1$ flips at rate $a$, and for $\psi(E) = E\,\mu(E)$ when the force is $E$, where the mobility $\mu(E)$ depends on the persistence and on the architecture of the unit cell.   The idea is that \eqref{1eff} summarizes effectively the joint influence of the persistence and obstacles but coarse grained to the translation invariant scale. Note that the homogenized mobility $\mu$ refers to the current in the positive $x-$direction and needs not be symmetric at all, as it summarizes the spatial escape rate and roughness need not be symmetric.\\
For computing the current we go to the Master equation for 
\eqref{1eff},
\begin{equation}
	\begin{cases}
		\partial_t \rho_+(x) = - \psi(c+\epsilon) \,\partial_x\rho_+(x) - a \left( \rho_+(x) - \rho_-(x) \right) \\
		\partial_t \rho_-(x) = - \psi(-c+\epsilon) \partial_x \rho_-(x) + a \left( \rho_+(x) - \rho_-(x) \right)
	\end{cases}
\end{equation}
where $\rho_{\pm}$ represents the density where $\sigma_t=\pm 1$.
Defining $\rho(x) = \rho_+(x) + \rho_-(x)$ for the total particle density and $m(x) = \rho_+(x) - \rho_-(x)$ for the chirality, we get
\begin{equation*}
	\begin{cases}
		\partial_t \rho(x) =-\frac 1{2} [\psi(c+\epsilon)-\psi(-c+\epsilon)]\, \partial_x m(x) - \frac 1{2}[\psi(c+\epsilon)+\psi(-c+\epsilon)]\, \partial_x \rho(x) \\
		\partial_t m(x) = -\frac 1{2}[\psi(c+\epsilon)-\psi(-c+\epsilon)]\, \partial_x \rho(x) - \frac 1{2}[\psi(c+\epsilon)+\psi(-c+\epsilon)]\,\partial_x m(x) - 2a\,m(x) 
	\end{cases}
\end{equation*}
On the other hand, writing $\psi_+ = \psi(c+\ve), \psi_- = \psi(-c+\ve)$,
\[
\partial^2_{tt} \rho = \frac 1{4}(\psi_+-\psi_-)^2\partial^2_{xx}\rho+ \frac 1{4}(\psi_+^2-\psi_-^2) \partial^2_{xx} m  + a(\psi_+-\psi_-) \partial_x m  - \frac1{2}(\psi_++\psi_-) \partial^2_{xt} \rho 
\]
and hence
\[
\partial^2_{tt} \rho + 2a \partial_t \rho =  \frac 1{4}(\psi_+-\psi_-)^2\partial^2_{xx}\rho+ \frac 1{4}(\psi_+^2-\psi_-^2) \partial^2_{xx} m   
- \frac1{2}(\psi_++\psi_-) \partial^2_{xt} \rho - a(\psi_++\psi_-)\partial_x\rho
\]
We still need
\begin{equation*}
	\partial^2_{xt} \rho = -\frac 1{2}(\psi_+-\psi_-)\partial^2_{xx} m - \frac 1{2}(\psi_++\psi_-) \partial^2_{xx} \rho	
\end{equation*}
to eliminate $\partial^2_{xx} m$, after which we find the modification of the telegraph equation,
\begin{equation}\label{mte}
	\partial^2_{tt} \rho + 2a \partial_t \rho + \frac 1{2} (\psi_++\psi_-)\partial^2_{xt} \rho =  -\psi_+\psi_-\,\partial^2_{xx}\rho- a(\psi_++\psi_-)\partial_x\rho\end{equation}
purely in terms of the particle density $\rho(x,t)$.  Note that when $\mu(E) = 1, \psi_+ = c+\ve,\psi_- = -c+\ve$ (no obstacles), the mobility does not depend on the force and the equations \eqref{1eff}--\eqref{mte} become
\begin{eqnarray}\label{mtes}
\frac{\id}{\id t}(x-\ve t) = c\;\sigma_t\nonumber\\
\partial^2_{tt} \rho + (2a+\ve\,\partial_x)\, (\partial_t \rho + \ve\partial_x\rho)&=&  c^2\,\partial^2_{xx}\rho\end{eqnarray}
(hyperbolic when $4(\ve^2-c^2)< \ve^2$).  Those do not yet take into account boundary conditions (interior of the channel).  Such one-dimensional models with bias and persistence have been studied analytically in \cite{pot}.  We remember from there that the current as well as the diffusion coefficient
are enhanced by persistence.\\
Continuing with \eqref{mte} we calculate the stationary current
\[
j = \lim_t \frac{\id}{\id t}\int\id x \,\rho_t(x)\, x
\]
by multiplying  \eqref{mte} with $x$ and integrating over $x$.
The mean velocity $v_t =  \frac{\id}{\id t}\int\id x \,\rho_t(x)\, x$ satisfies
\[
\partial_t v_t + 2av_t = a(\psi_++\psi_-),\;\quad\;
v_t = e^{-2at}v_0 +\frac{\psi_++\psi_-}{2}(1-e^{-2at})
\]
which asymptotically in time $t$ becomes
\begin{eqnarray}\label{jj}
v_t \rightarrow j &=& \frac{\psi(c+\ve)+\psi(-c+\ve)}{2} \nonumber\\&=& \frac{c}{2}\,[\mu(c+\ve)- \mu(-c+\ve)] + \frac{\ve}{2}\,[\mu(c+\ve)+ \mu(-c+\ve)]
\end{eqnarray}
The current-charateristic $j = j(\ve; c,a)$ is thus largely decided by the behavior of the effective mobility, which also depends on the persistence $a$ (and on the local roughness).  There are in fact two other length scales besides $L$, which we can take as the size of the unit cell or obstacles:  there is $\ve/a$ which is the bias length and $c/a$ which is the persistence length.\\
Let us first take $c\gg\ve$. We see that $j>0$ unless $\mu(\ve-c)$ is large enough.  For example, for zero external field $\ve =0$, $\mu(-c) > \mu(c)$ yields a negative current $j(\ve=0)<0$ at zero field, which is very well possible if escape opposite to the field becomes easier than with the field. Larger persistence is then expected to increase the difference $\mu(-c) - \mu(c)$ which lowers the current at small bias.\\  For large $\ve\gg c$ we have $j\simeq \ve\mu(\ve)$, and the decay of the mobility in $\ve$ decides the characteristic for large pumping.  We may expect that if there is a horizon of unblocked escape, larger persistence (in that direction) increases $\mu(\ve)$, and more for larger $c$.\\

We see next more specifically for a periodic channel with obstacles how the predictions of the above effective analysis get realized, in particular when the unit cell is asymmetric.

\section{Model with obstacles}\label{mod}

\begin{figure}[h]
\centering
\subfigure[]{\includegraphics[width=0.6\textwidth]{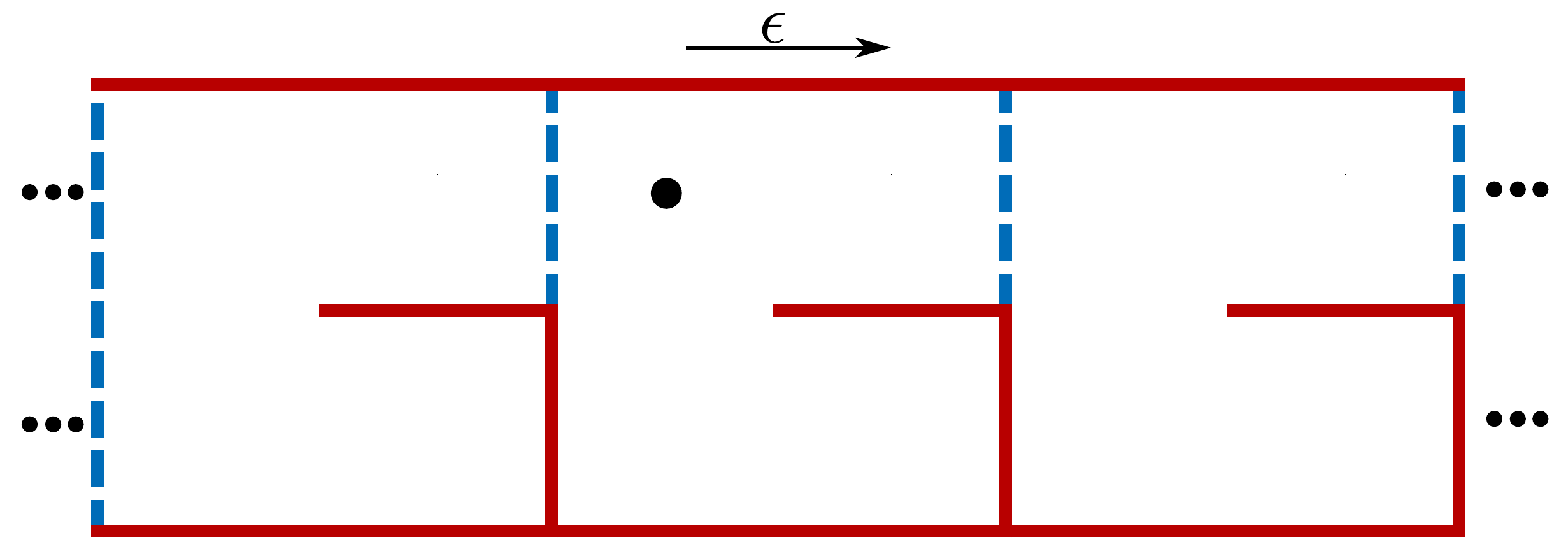}}
\subfigure[]{\includegraphics[width=0.6\textwidth]{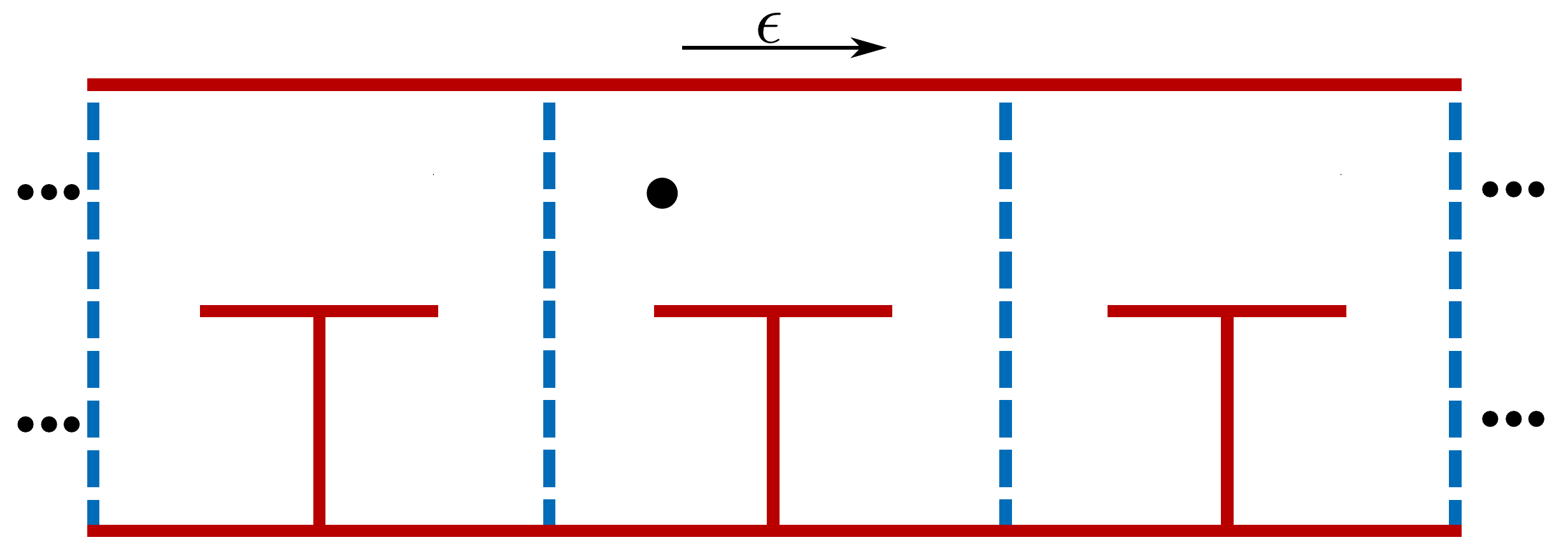}}
\caption{Periodic channel with obstacles in the form of hooks \textbf{(a)} asymmetric hook and \textbf{(b)} symmetric hook. Solid lines (red) are obstacles, the dashed lines (blue) only indicate the boundaries of the unit cells.} 
\label{fig: setups}
\end{figure}

We restrict ourselves to two types of obstacles, both time-symmetric but one which is spatially asymmetric (with one type of hook) and one which is spatially symmetric.  The space is quasi-one dimensional, either continuous, in Section \ref{conts}, or discrete, in Section \ref{dis}.  See  Fig.~\ref{fig: setups} for the setup of the rough channel, showing three unit cells.  We have also indicated the direction of the nonconservative external field of strength $\ve$. 

 In the numerical simulations, time is discrete with steps of size $\Delta t$. The current $j$ is measured during an interval $\left[0, t_\text{max} \right]$ after the system had time to relax. The stationary current is then obtained as
\begin{equation}\label{eq: current}
	j = \frac{\phi_r - \phi_\ell}{t_\text{max}}	
\end{equation}
where $\phi_r$ and $\phi_\ell$ are the number of particles leaving the unit cell through the left,  respectively  the right boundary of the unit cell. Starting positions of $N$ particles are  chosen randomly within the unit cell and evolve independently.

\subsection{Continuum model}\label{conts}

The motion \eqref{eq: RTP in field} is supplemented with boundary conditions.  We refer to 
Figs.~\ref{fig: setups} for the basic architectures that are repeated periodically. 
For the obstacles (solid lines in the figures) the boundary conditions are hovering, meaning that when a particle hits these boundaries it will continue by keeping (only) its velocity component parallel to the surface.  Particles cannot move through obstacles. For the rest, when not at an obstacle, boundary conditions are periodic in the (horizontal) $x-$direction: if a particle leaves the unit cell at these boundaries it enters the unit cell at the opposing side.\\



The simulations use a unit cell with sides of unit length, the walls of the obstacles have half the length of the cell making them span over one-fourth of the area. Other parameters are $\Delta t = 0.001$, $t_\text{max}= 500$, $N=10000$ and $c=1$ unless stated otherwise.


\subsection{Discrete model}\label{dis}

\begin{figure}[h]
	\centering
	\includegraphics[width=0.6\textwidth]{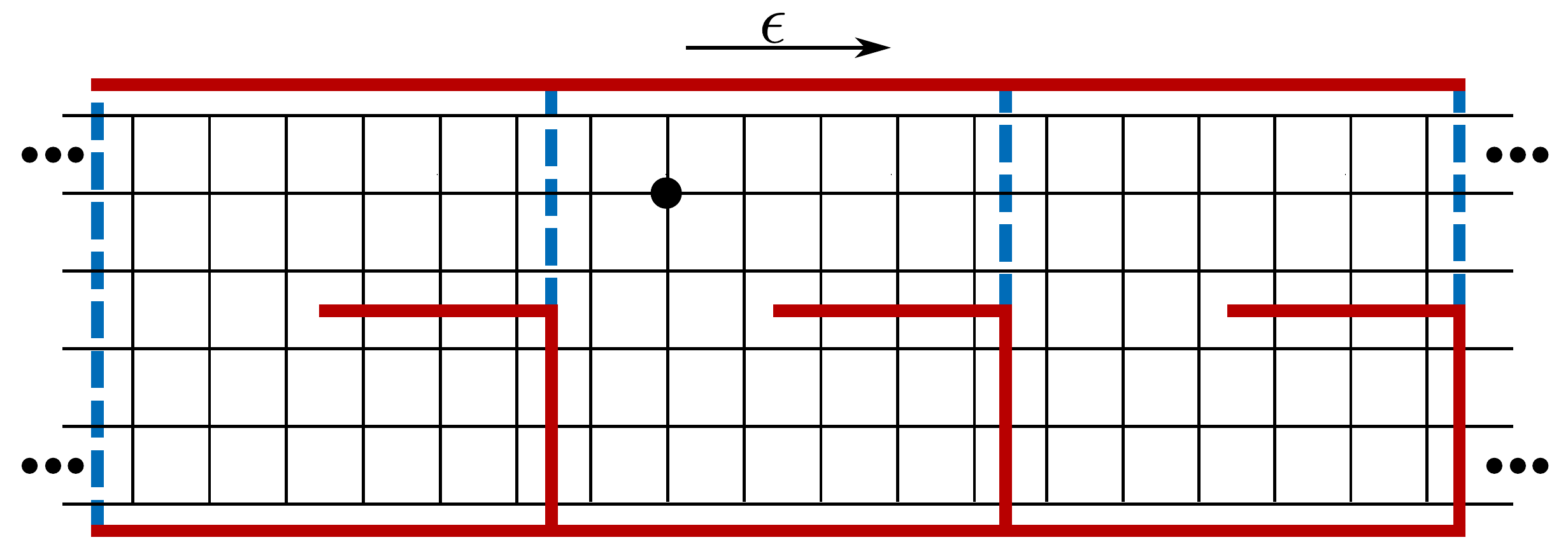}
	\caption{Setup for the discrete lattice model with a hook at the right. The same is done for the symmetric case of Fig.~\ref{fig: setups}(b).}
	\label{fig: obstacles discrete left}
\end{figure}

For run-and-tumble motion on the discrete lattice, we deal with a random walk showing persistence and subject to an external bias \cite{revi,pot,pat}.  Hopping is between lattice sites except that particles cannot cross the walls (the solid (red) lines in Fig.~\ref{fig: obstacles discrete left}).  The unit cell is again repeated periodically.   Study of the current-characteristic for active particles on a discrete lattice with obstacles is new and requires some new elements to take care of.\\

The run-and-tumbling follows the same idea as in the continuum case but combining persistence with bias in a continuous time random walk in two dimensions requires more explanations.  The (internal) spin direction also becomes discrete and can either point right, left, up or down. The tumbling is still uniformly to one of the other directions with a rate $a$, i.e., with uniform and symmetric transition rates $k($right,left$)=k($right,up$) = k($right,down$)=\ldots=a$.  The particle is persistent in one of the four directions and is subject also to an external field $\ve$ pointing to the right in the (horizontal) $x-$direction.   We now specify the hopping rates.  There will be each time four hopping rates, and each is set to zero when an obstacle prevents the hopping (see again Fig.~\ref{fig: obstacles discrete left}). The hopping rates depend on the current direction of persistence (with amplitude $c$) and the external field $\ve$.\\

To better understand the logic of the formul{\ae}  that follow, we first consider a one-dimensional integer lattice with sites $j$, for modelling a run-and-tumble particle in an external field (without obstacles).  For that purpose we propose the following rates.
\begin{figure}[h]
	\centering
		\includegraphics[width=\textwidth]{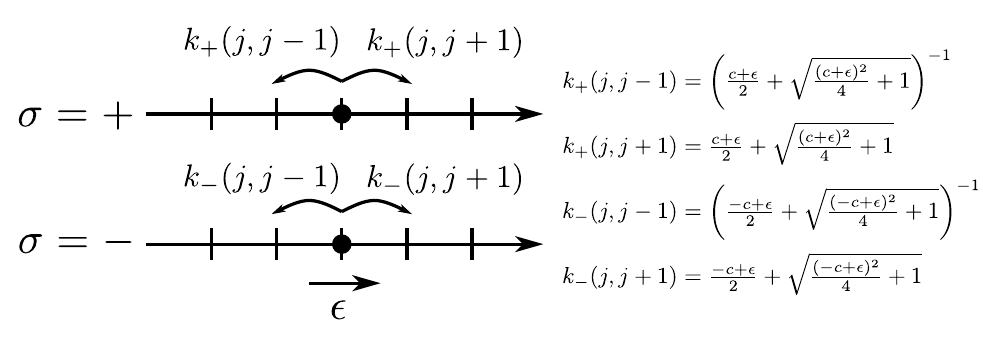}
	\caption{Setup for a one-dimensional RTP in a field $\epsilon$.}
	\label{fig: RTP in field 1d}
\end{figure}

When the direction of persistence is to the right (subscript $+$), we take rates
\[k_+(j,j+1) = \exp b_+,\qquad k_+(j,j-1) = \exp -b_+  
\]
and similarly
\[k_-(j,j+1) = \exp b_-,\qquad k_-(j,j-1) = \exp -b_-
\]
for the hopping rates when the persistence is to the left (with subscript $-$).  The master equation for the density $\rho_+(j)$, respectively $\rho_-(j)$ is
\begin{equation}
\begin{cases}
\partial_t \rho_+(j) =  e^{b_+} \left( \rho_+(j-1) -\rho_+(j) \right) +  e^{-b_+}\left( \rho_+(j+1)-\rho_+(j) \right) - a \left( \rho_+(j) -\rho_-(j) \right)\\
\partial_t \rho_-(j) =  e^{b_-} \left( \rho_-(j-1) -\rho_-(j) \right) +  e^{-b_-}\left( \rho_-(j+1)-\rho_-(j) \right) - a \left( \rho_-(j) -\rho_+(j) \right)
\end{cases}
\end{equation}

Clearly then, we should take $2\sinh b_+ = c+\ve$ and $2\sinh b_- = -c + \ve$ for mimicking the run-and-tumble motion $\dot x_t = c\sigma_t + \ve$ on $\mathbb{R}$, with $\sigma_t=
\pm 1$ tumbling at rate $a$.  Observe that on the lattice, particles are still allowed to go against the direction of persistence in the same way as they can go against the external field.  In the continuum limit (scaling time and space in the same way) we obtain the correct run-and-tumble dynamics.  There exist other versions of lattice dynamics of run-and-tumble particles, in particular for $\ve=0$ where we can simply take one-way rates depending on the direction of persistence; see for example \cite{ha,pot}.  However, in our model, we can easily put $c=0$ and keep the motion of a biased random walker for $\ve \neq 0$.\\

We extend the above idea to the two-dimensional lattice, e.g. in Fig.\ref{fig: obstacles discrete left}.  It is important that when the external field is zero $\ve=0$ the hopping rates in the three other directions from the direction of persistence are just equal.  We now put subscripts $r,\ell, u, d$ for  the direction of persistence, indicating right, left, up or down respectively.  We then denote $p=r,\ell, u, d$ for the persistence direction, and get the hopping rates
\[k_p(j,j +\boldsymbol{\hat{e}_x}) = e^{\alpha_p},\,\quad k_p(j,j -\boldsymbol{\hat{e}_x}) = e^{\beta_p},
\;k_p(j,j +\boldsymbol{\hat{e}_y}) = e^{\gamma_p},\,\quad k_p(j,j -\boldsymbol{\hat{e}_y}) = e^{\lambda_p}
\]
with
\begin{equation*}
\begin{cases}
\sinh\alpha_r = \frac{c+\epsilon}{2} \\
\sinh\beta_r = -\frac{c+\epsilon}{2} \\
\sinh\gamma_r = \frac{-c}{2} \\
\sinh\lambda_r = \frac{-c}{2}
\end{cases},
\begin{cases}
\sinh\alpha_\ell = \frac{-c+\epsilon}{2} \\
\sinh\beta_\ell = -\frac{-c+\epsilon}{2} \\
\sinh\gamma_\ell = \frac{-c}{2} \\
\sinh\lambda_\ell = \frac{-c}{2}
\end{cases},
\begin{cases}
\sinh\alpha_u = \frac{-c+\epsilon}{2} \\
\sinh\beta_u = -\frac{c+\epsilon}{2} \\
\sinh\gamma_u = \frac{c}{2} \\
\sinh\lambda_u = \frac{-c}{2}
\end{cases},
\begin{cases}
\sinh\alpha_d = \frac{-c+\epsilon}{2} \\
\sinh\beta_d = -\frac{c+\epsilon}{2} \\
\sinh\gamma_d = \frac{-c}{2} \\
\sinh\lambda_d = \frac{c}{2}
\end{cases}
\end{equation*}
explicitly in terms of the physical amplitudes $c$ and $\ve$.  The above may seem complicated updating formul{\ae} but we claim they are the physically correct ones for RTP in an external field on a 2dimensional lattice.\\

Again the unit cell has sides of unit length, there are 10 rows and 10 columns of lattice sites, leading to an inter site distance $\delta_x = \delta_y = 0.1$. The walls are placed between lattice points prohibiting hopping over them. The asymmetric hooks consist of a barrier between columns 10 and 1 for the 5 bottom rows of the lattice and one in columns 5 to 10 between rows 5 and 6. The symmetric hooks, on the other hand, have one barrier between columns 5 and 6 for rows 1 to 5 and a second one above columns 3 to 8 between row 5 and 6. The other parameters used in the simulations are $\Delta t = 0.01$, $t_\text{max}=500$, $N=10000$ and $c=0.5 \text{ or } 1$ unless mentioned otherwise.
\begin{figure}[h]
	\centering
	\subfigure[]{\includegraphics[width=0.49\textwidth]{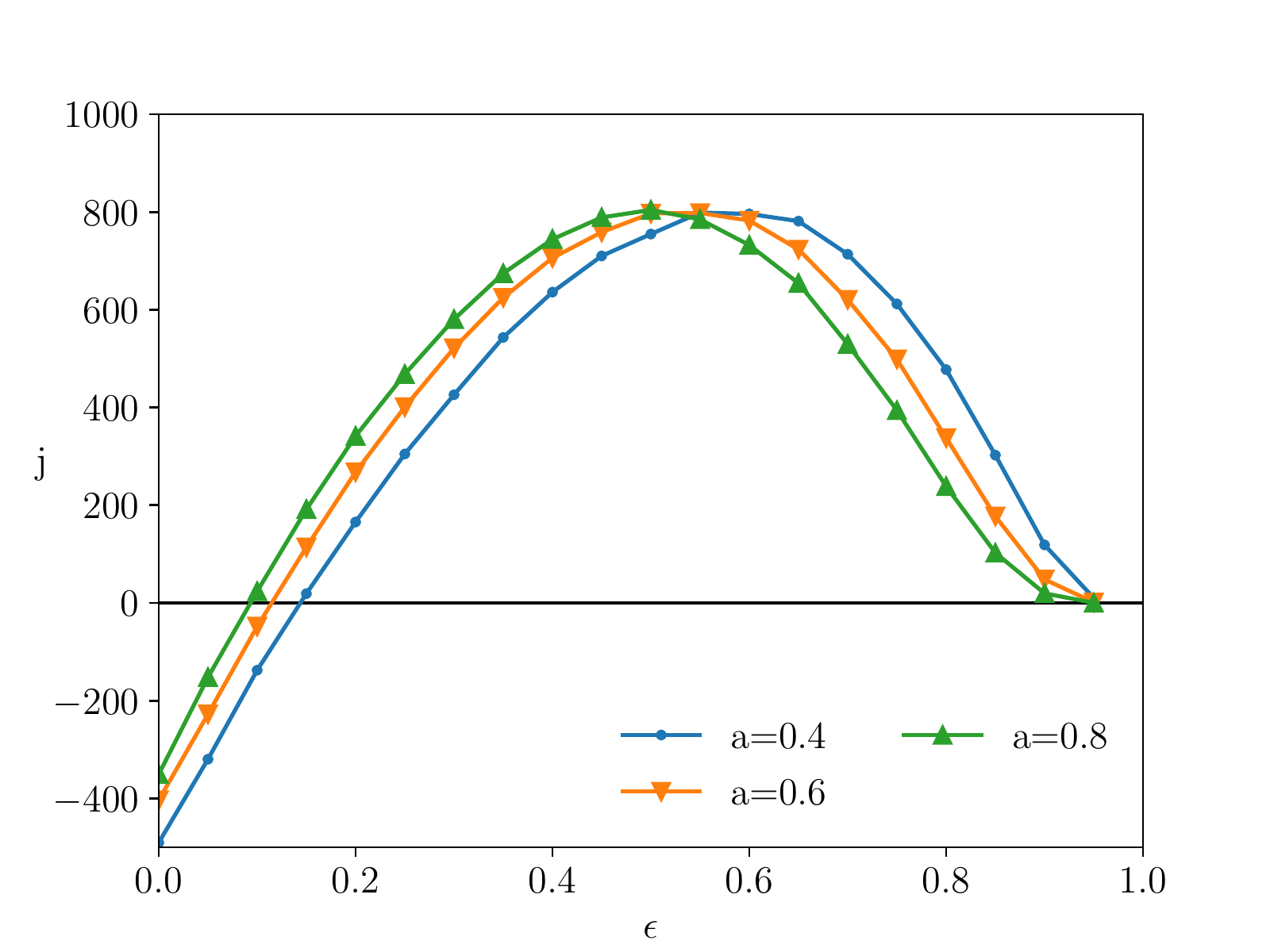}}
	\subfigure[]{\includegraphics[width=0.49\textwidth]{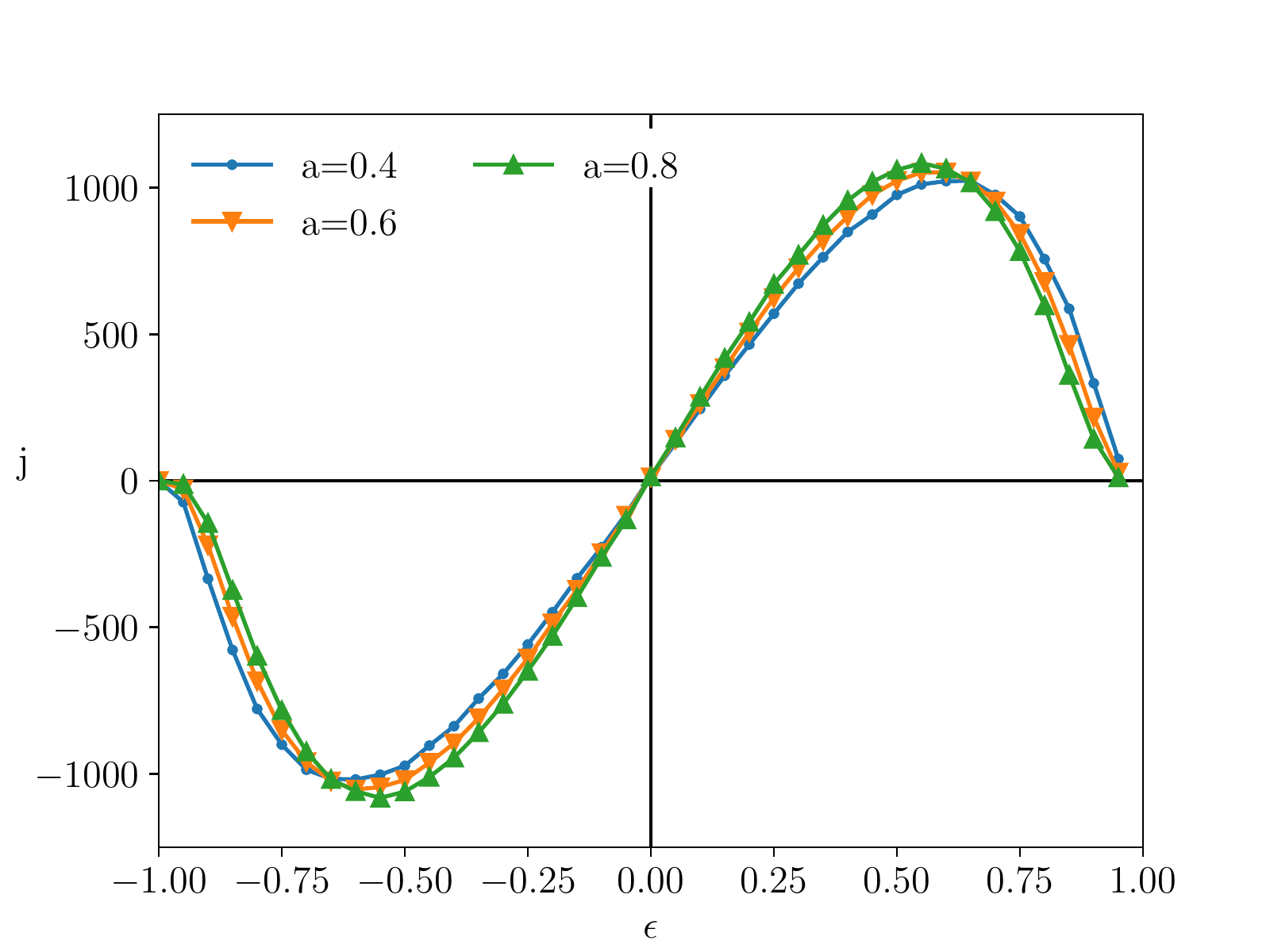}}
	\caption{The current-field characteristic for different persistence values in the continuum with $c=1$.  Left: asymmetric hook as in Fig.~\ref{fig: setups}(a).  Note the current is negative for zero field, and still lower for higher persistence at low bias, while the current increases with the bias at larger field. Right: symmetric hook as in Fig.~\ref{fig: setups}(b).  The current depends less on the persistence at small field values, but still increases with persistence at large fields.}
	\label{fig: results different a continuum}
\end{figure}

\begin{figure}[h]
	\centering
	\subfigure[]{\includegraphics[width=0.49\textwidth]{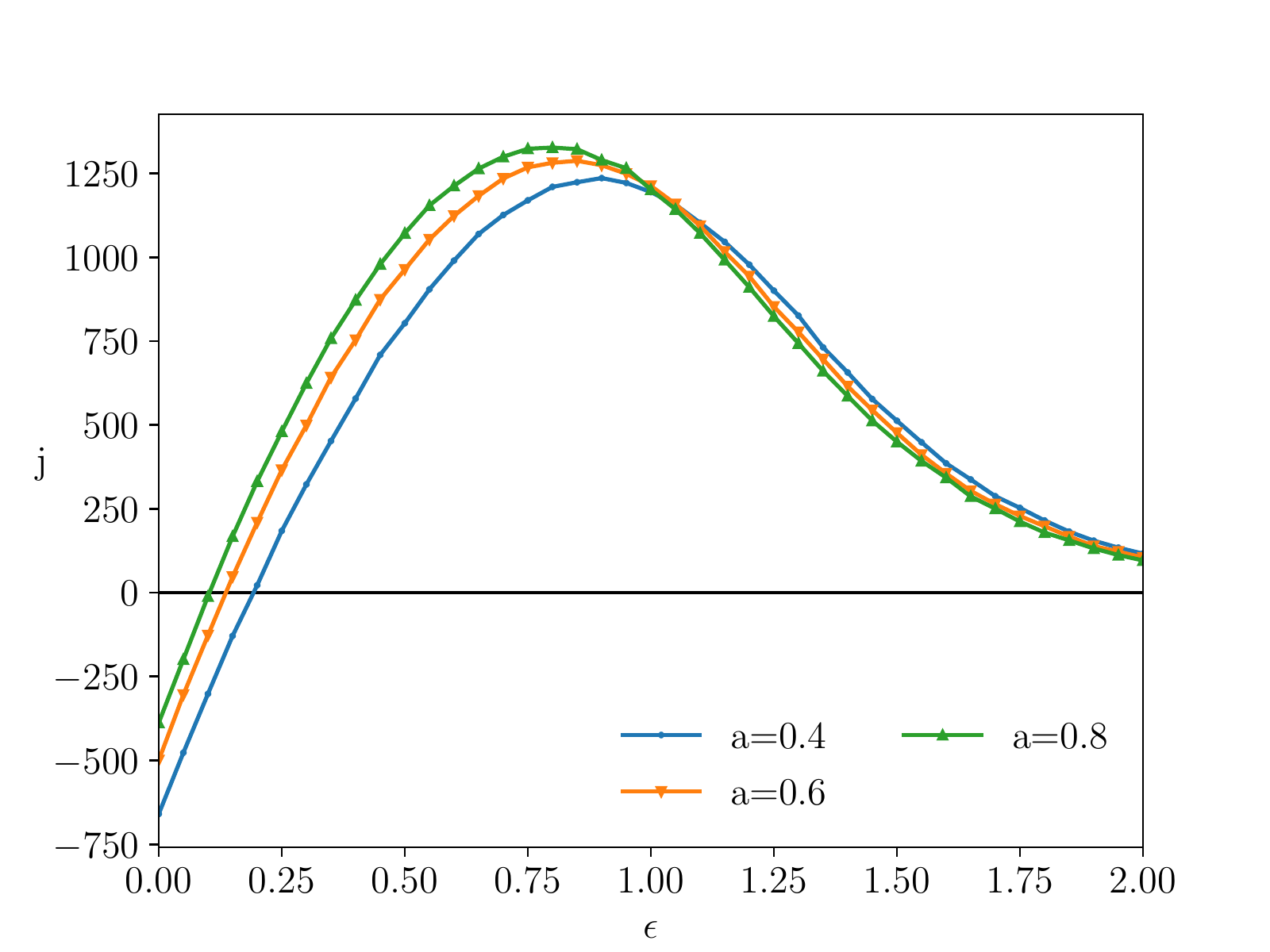}}
\subfigure[]{\includegraphics[width=0.49\textwidth]{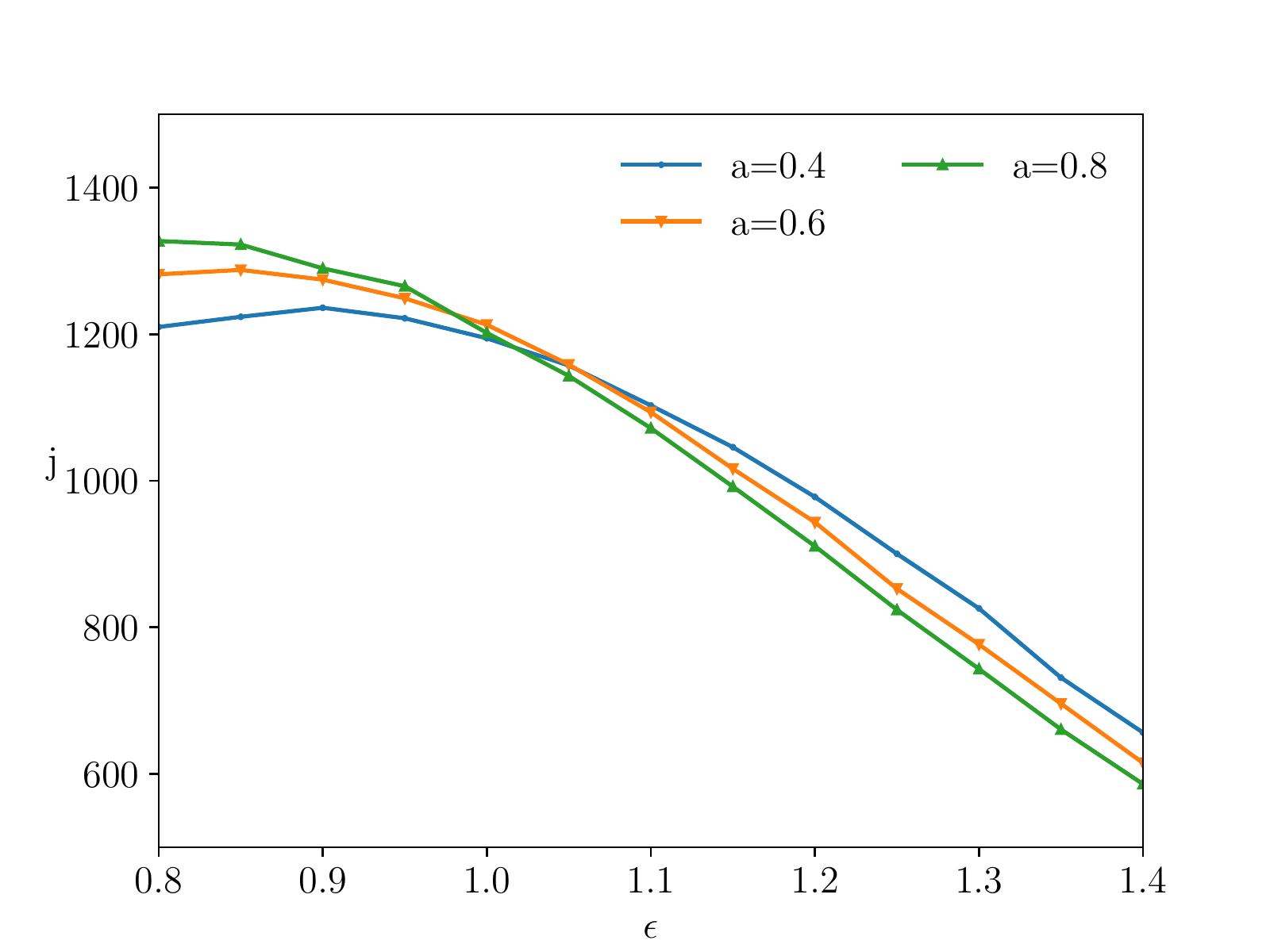}}
	\subfigure[]{\includegraphics[width=0.49\textwidth]{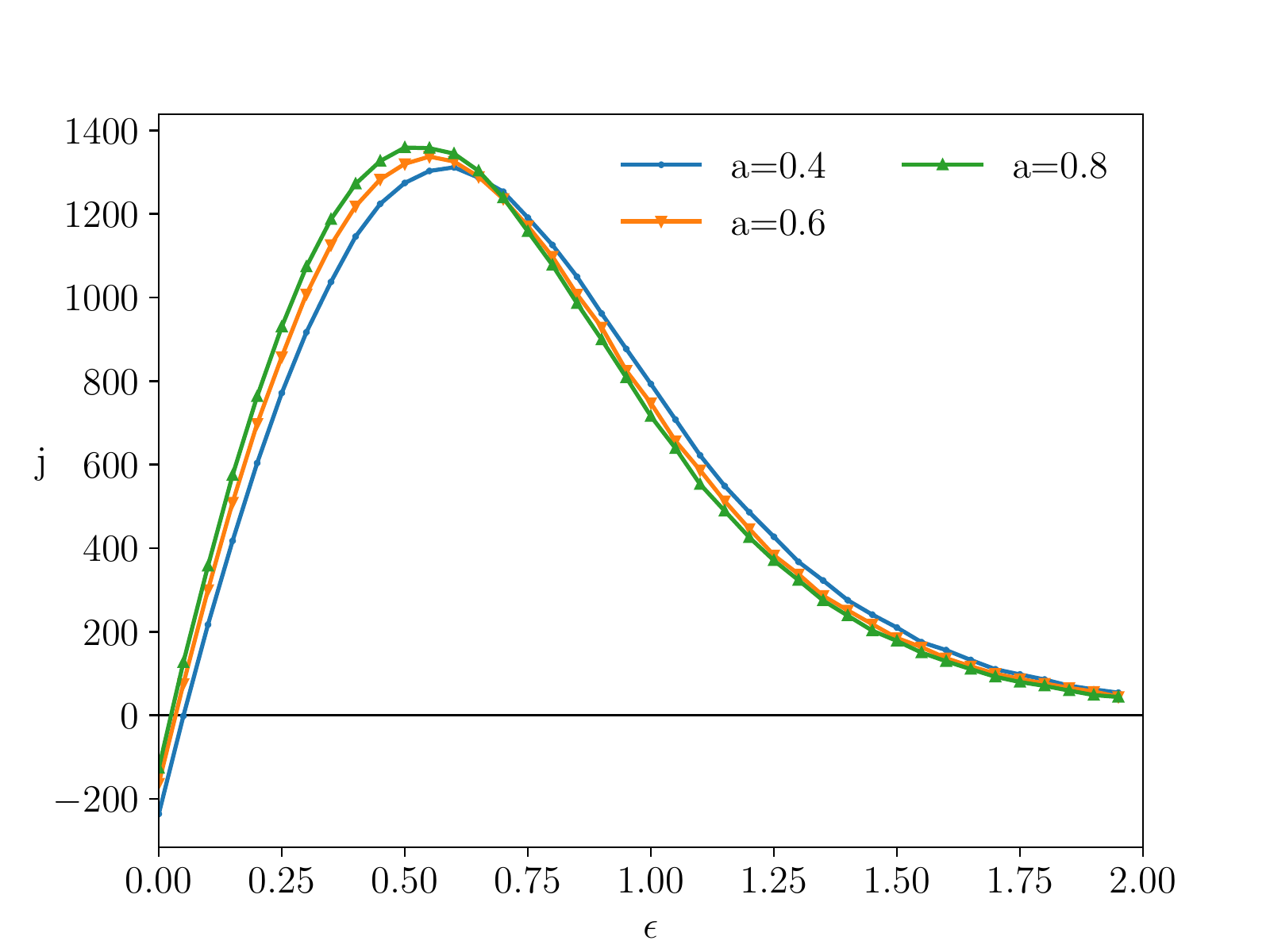}}
\subfigure[]{\includegraphics[width=0.49\textwidth]{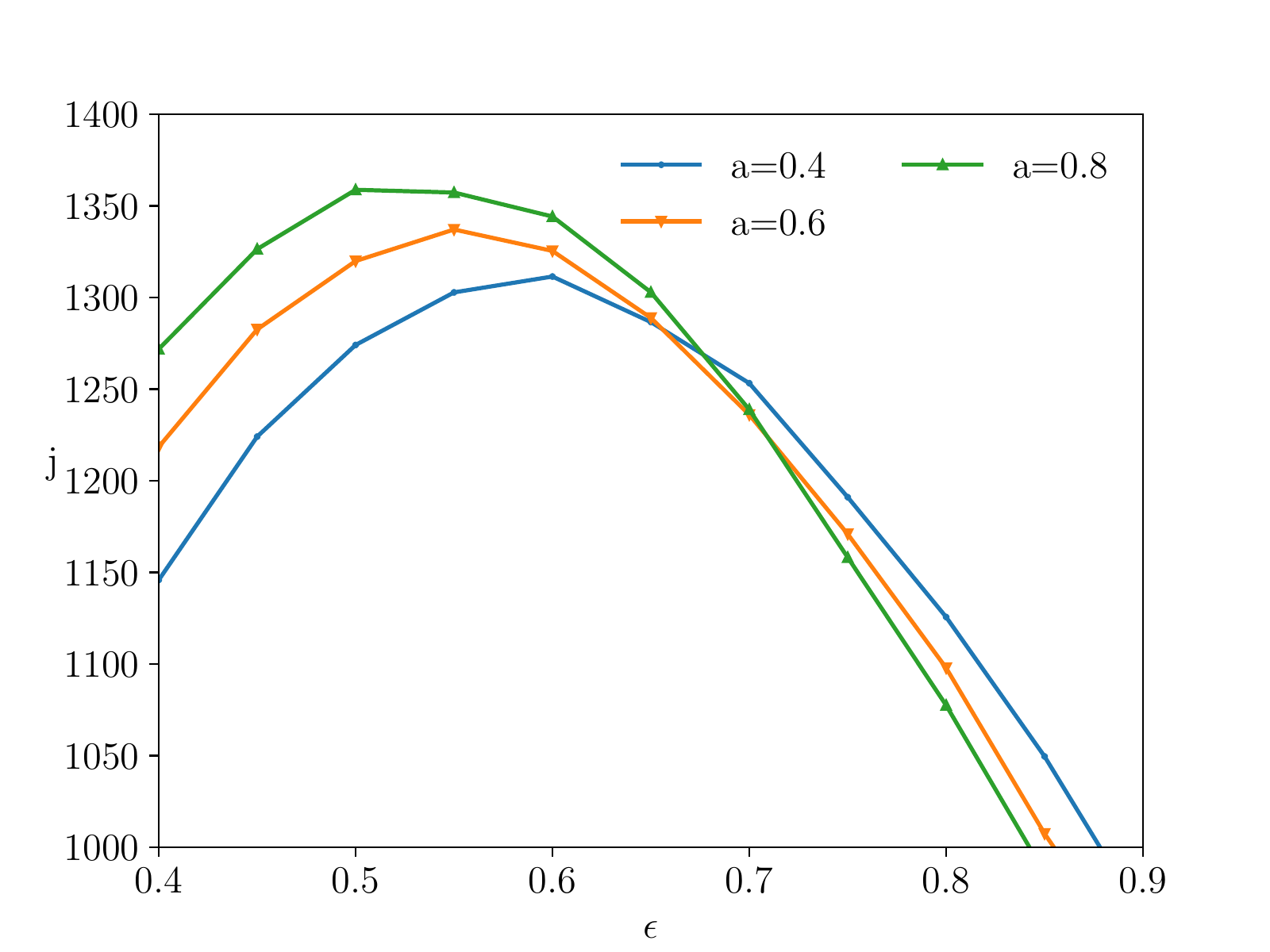}}
	\caption{The current-field characteristic for different persistence values for the asymmetric hook as in Fig.~\ref{fig: setups}(a). $c =1$ for (a) and (b) and $c =0.5$ for (c) and (d).  Left:  The current characteristic is similar to the continuum Fig.~\ref{fig: results different a continuum}(a), except in the large field regime where persistence matters less. Right:  The current switches from decreasing to increasing with the persistence time $a^{-1}$, around some value $\ve_c$.}
	\label{fig: results different a discrete}
\end{figure}

\section{Results}\label{res}

The current characteristic gives the dependence of the stationary current on the external field $\ve$, on the persistence time $a^{-1}$ and on the coupling amplitude $c$.\\
Our findings are summarized in Figs.~\ref{fig: results different a continuum} for the continuum model and in Figs.~\ref{fig: results different a discrete} for the discrete case.  To the left, we see the case for an asymmetric trap as in Fig.~\ref{fig: setups}(a) and to the right for a spatially symmetric obstacle as in Fig.~\ref{fig: setups}(b).

We find the following behavior (in the continuum and on discrete lattice):
\begin{enumerate}
	\item For small enough (including zero) external field and for asymmetric trapping there appears a current opposite the external field. In fact, for zero field and in a closed box with asymmetric trapping there appears a density gradient.
	\item There exists a threshold value $\ve_c>0$ so that for $0\leq \ve < \ve_c$ the current decreases  with persistence time $a^{-1}$, and for $\ve_c < \ve$ the current increases with the persistence. This $\ve_c$ depends on the coupling amplitude $c$ and occurs around $\ve = O(c)$.\\
		For example, compared to the passive case, the current in the active case is first smaller and later gets larger.  The dying of the current is postponed longer for greater activity.
	\item The effect of persistence becomes more important for larger speed $c$.  For fixed persistence, the current grows with $c$ but the maximal current is not monotone in $c$.
\end{enumerate}

%

\begin{figure}[h]
	\centering
	\subfigure[]{\includegraphics[width=0.49\textwidth]{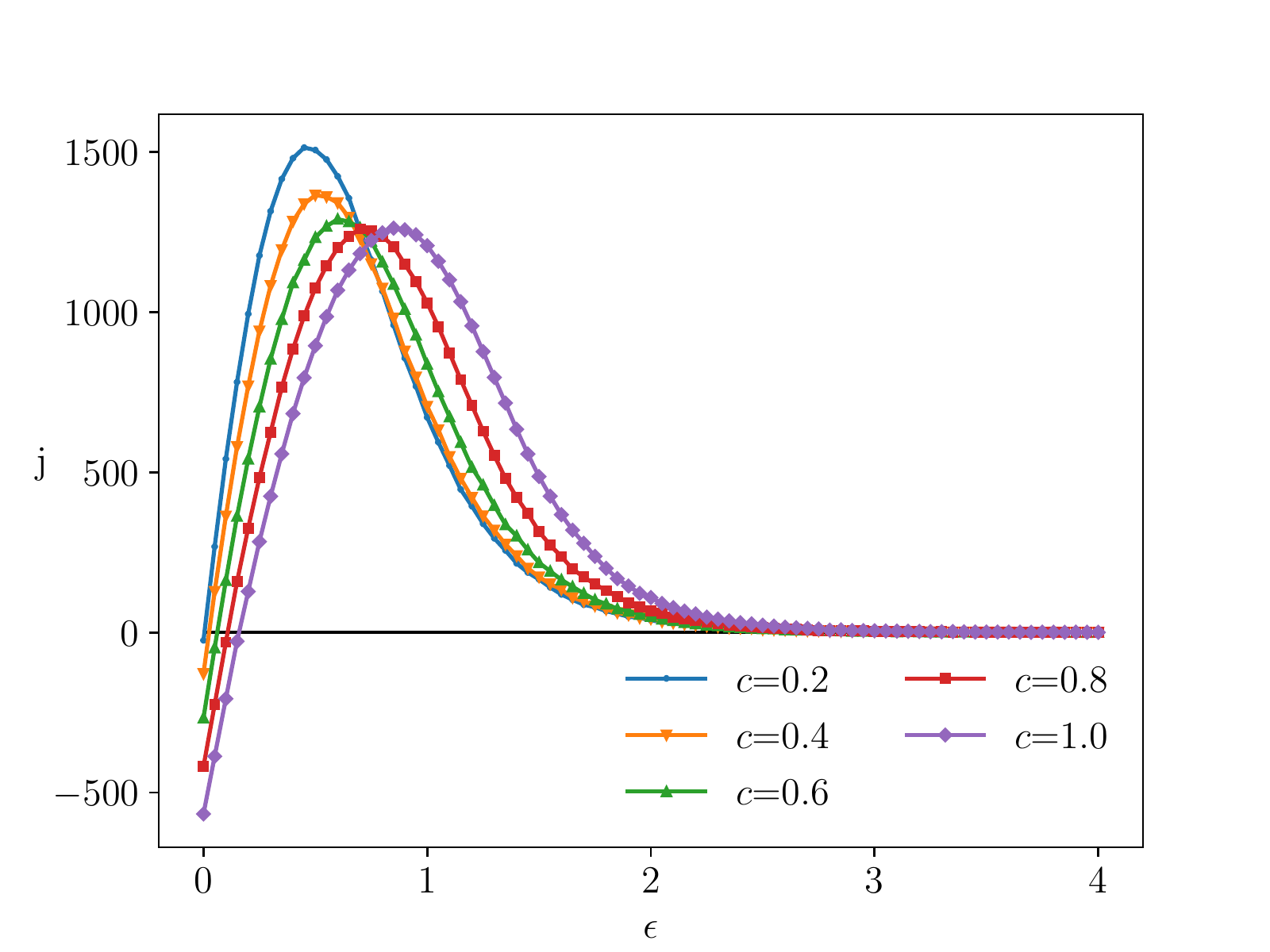}}
	\subfigure[]{\includegraphics[width=0.49\textwidth]{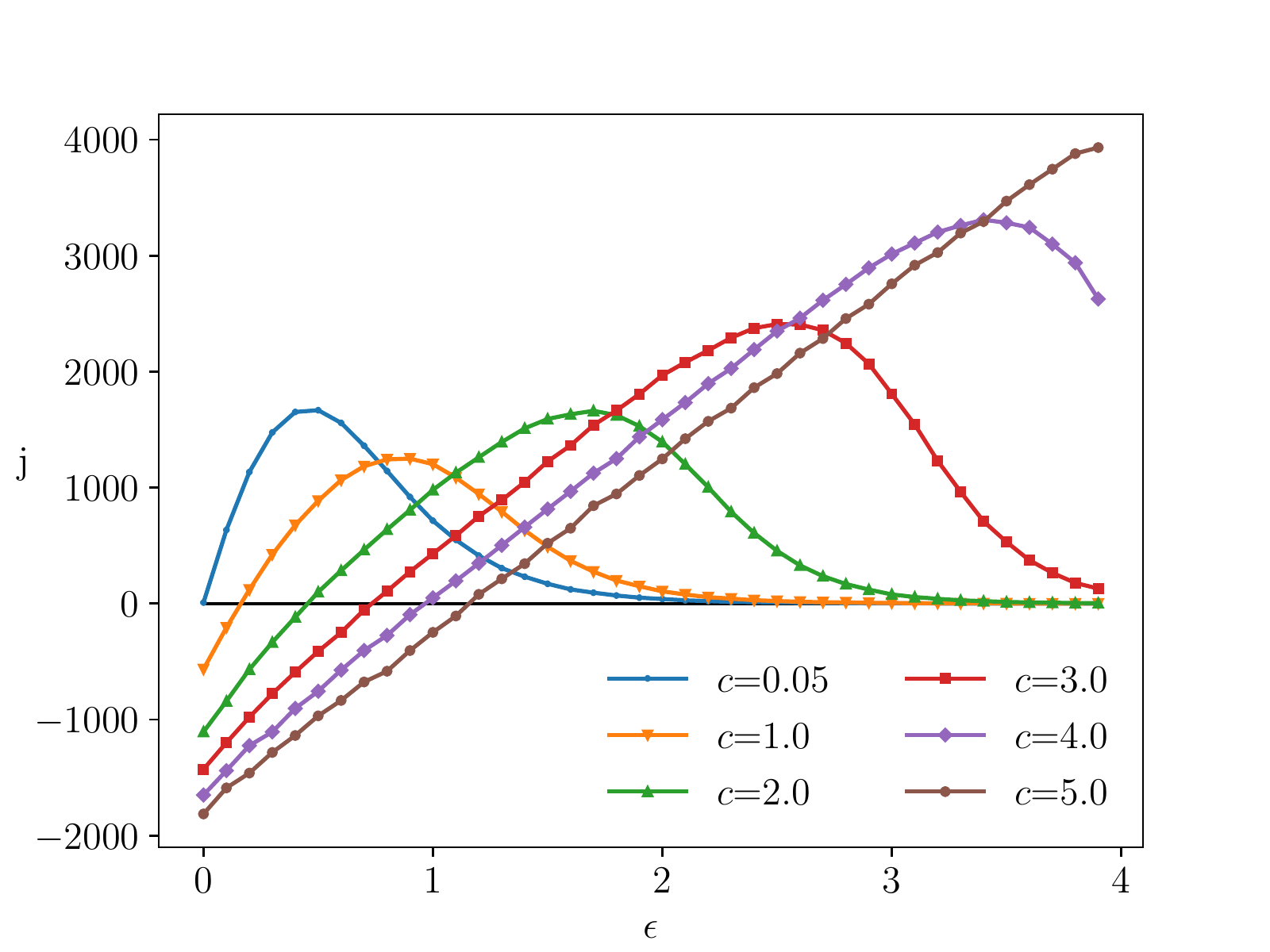}}
	\caption{The current-field characteristic for different coupling amplitude $c$ at $a=0.5$ on the lattice. The field at which the current is maximal increases with $c$ and the maximum value is smallest for $c =1$.}
	\label{fig:}
\end{figure}
\begin{figure}[h]
	\centering
	\includegraphics[width=0.49\textwidth]{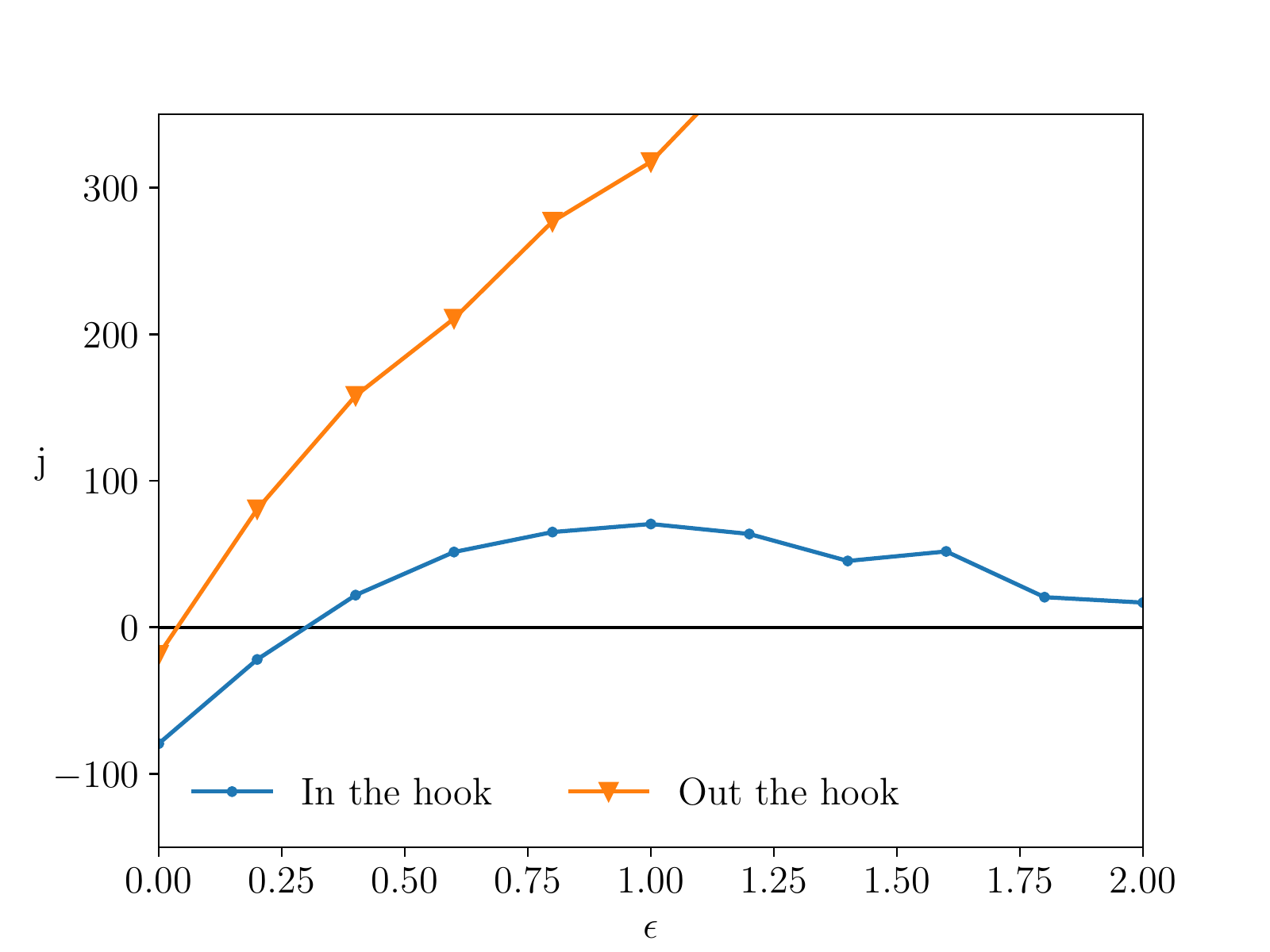}
	\caption{RTP trapped in the right-hook of Fig.~\ref{fig: obstacles discrete left} typically leave the unit cell to the left, especially at high persistence (here: $a=0.5$ and $c=1$). To evaluate the effect of the starting position, the current \eqref{eq: current} is only measured until $t_\text{max}= 10$.}
	\label{fig: discussion discrete}
\end{figure}

\section{Discussion}\label{theo}
We refer to our findings in the previous section, and we discuss how they are clarified from the effective model in Section \ref{theor}.
\begin{enumerate}
	\item Asymmetric traps yield a nonzero current in zero fields.  The active system is clearly not in equilibrium even at $\ve =0$.  The current generation (opposite to the trap) is similar to what happens in the case of the Parrondo paradox or for flashing ratchets where the nonequilibrium condition conspires with asymmetry \cite{parr,han,cub}.  It was also found recently in \cite{rei2} for active particles. Here the point is foremost that particles starting inside the trap typically contribute to the current in the opposite direction; see Fig.~\ref{fig: discussion discrete}. Particles in the trap tend to leave the unit cell to the left, as they need persistence in a direction oppositie the wall, ending up in the left half of the cell. The analytical analogue is that $\mu(-c)> \mu(c)$ in \eqref{jj} and as announced there.
	\item  The initial decrease in the small-field current (including $j_{\epsilon=0}$) is related to the previous remark, as for small field and higher persistence, there is a substantial contribution of negative flux due to particles that are trapped. For higher persistence, the particles may go for a longer time against the field and it thus takes a larger field to fully trap them.  As $a \to \infty$ the particle becomes passive and $j_{\epsilon=0}=0$.
\item  For large external field $\ve$, the effective field $\ve \pm c$ is always positive and trapping occurs finally inside the hook.  The current goes down exponentially fast in $\ve$ as the mobility $\mu(E)$ is exponentially small in $E$ as for the passive case \cite{neg}.
What we see here in general is nonequilibrium response where a positive correlation between current and frenesy in the {\it original} dynamics produces a negative contribution; see e.g. \cite{resp}. From the results in the present paper we conclude that persistence in self-propulsion is able to postpone the trapping-effect.  At large enough pumping, RTP obtain an increased current for higher activity. Refinements where the tumbling rate depends on the neighborhood of an obstacle or on the direction of the external field \cite{co} can be considered in further studies.
\end{enumerate}
\vspace{1cm}
\noindent {\bf Acknowledgment:}
Bram Bijnens is grateful for the opportunity of the student internship at the Institute for Theoretical Physics in Leuven, during which the present study was completed.


\end{document}